**Standardisation of work: co-constructed practice**

Gunnar Ellingsen

Department of Telemedicine, University of Tromsø, 9000 Tromsø, gunnar.ellingsen@unn.no

Eric Monteiro

Department of Computer and Information Systems, Norwegian University of Science and Technology, 7491 Trondheim, eric.monteiro@idi.ntnu.no

Glenn Munkvold

Department of Information Technology, Nord-Trøndelag University College, 7729 Steinkjer, glenn.munkvold@idi.ntnu.no

# Standardisation of work: co-constructed practice

# Abstract

There is strong pressure to achieve greater uniformity, standardisation and application of best practices in the service professions, a sector which is growing in presence and importance. At the same time, there is a conflicting demand for the delivery of high-quality (or highly priced or 'knowledge intensive') specialised or localised services. Our paper analyses information systems' embedded efforts of standardising service work through an in-depth interpretative study of an ongoing standardisation initiative within the field of nursing. Nursing provides a graphic illustration of the dilemmas involved in the standardisation of service work. In nursing, standardisation is commonly a feature of projects to improve both efficiency and quality in health care. In contrast to the dominant conception of standardisation as a largely top-down, imposed process, we offer a view of standardisation as incomplete, co-constructed with users and with significant unintended consequences. The paper contributes by i) developing a theoretical perspective for the standardisation of information-system-embedded service work and ii) operational and practical implications for system design and health care management.

# 1   Introduction

Today, corporate and public-sector entities are under strong and growing pressure to respond to inherently conflicting concerns: on the one hand, achieving economies of scale through the dissemination of 'best practices' and of standardised routines and procedures; on the other hand, an increasing demand for individualised services and products (Brunsson & Jacobsson, 2000; Bowker & Star, 1999**;** de Guy & Salaman, 1992). This dilemma is especially acute in the field of service delivery, as Leidner (1993) reminds us, because high-quality services are characterised specifically by the perception that they are  not standardised but that they are sensitive to specific customers' needs (Alvesson**,** 2001).

Given the deeply embedded role of information systems (IS) in the ongoing transformation of modern organisations, we explore the effects of the conflicting forces involved in standardising service work in practice, analytically as well as operationally.

Standardisation within information systems is not new. There is a long history of formal or de jure standardisation of programming languages, communication protocols and exchange formats (Schmidt & Werle, 1998). There is an even stronger tradition of de facto standards for applications, operating systems and file formats (Kahin & Abbate 1995; Hanseth et al., 1996). What has received considerably less attention in IS research, however, is the study of IS-based initiatives for the standardisation of work and routines (Timmermans & Berg, 1997; Bowker et al., 1995). Given the growing presence and importance of the service sector, it is vital that IS research extends its focus from the standardisation of artefacts and products to include standardised, IS-embedded service work as well.

Our paper is based on an in-depth interpretative study of an ongoing IS-based intervention for the standardisation of one type of service work, namely nursing. The relevance of our case – the standardisation of planning, documentation and delivery of care for elderly, psychiatric patients at one ward at the University Hospital in Northern Norway – is associated with characteristic aspects of the case. Modern nursing is embedded in a highly *politicised* and *institutionalised* arena where governmental and managerial rules, regulations and policies are negotiated against local concerns and priorities. The need to curb large and seemingly ever-increasing health care *expenditure* is an explicit feature of managerial agendas for the increased standardisation of health care work (Timmermans & Berg, 2003). In contrast to many wards in Western hospitals, the ward which we selected for our study has a strong presence of *interdisciplinary* work. The effective treatment of elderly psychiatric patients cuts across disciplinary boundaries between nursing, medicine and physiotherapy. From the perspective of knowledge sharing, nursing provides a good illustration of the tensions and trade-offs between narrative forms of knowledge and (efforts of) codified forms (Bruner, 1990; Boland & Tenkasi, 1995; Orr, 1996).

A key feature of using IS as a means to standardise service work lies in the challenge – manifest in health care delivery – to aim for efficiency and productivity gains as well as for improvements in quality *simultaneously*. Our intention is to develop an understanding of information systems embedded in the standardisation of service work. More specifically, we critically discuss prevailing approaches portraying standardisation as an "iron grid" imposed from the top down which subjects merely have to comply with (Schmidt & Werle, 1998). We outline an alternative, transformative perspective on standardisation as incomplete, co-constructed with users and with significant unintended consequences. A particularly interesting aspect of this is the way in which the same effects tend to emerge in a wide range of settings, i.e. for other users, in other circumstances and in other locations. Furthermore, we contribute by highlighting practical operational, implications for IS design and management derived from our conceptualisation of standardisation.

In the rest of this paper, section 2 describes in more detail some experiences and conceptualisations of standardisation in health care. Standardisation initiatives relating to nursing are discussed in particular. Section 2 elaborates on our transformative, co-constructive perspective on standardisation of service work, which forms the conceptual core of our paper. In section 3 we describe and reflect on methodological issues around our in-depth case study of standardising work at a ward at the University Hospital of Northern Norway during the implementation of IS-based nursing plans. Section 4 contains a chronological case narrative, describing the background, process and perceptions associated with introducing and using nursing plans to standardise documentation and content of work. Section 5 contains an analysis, and is divided into four parts. The first three parts discuss degrees of deviations from intended use of the standardised plan: from smaller adjustments and tinkering to more radical transformations which warrant our label of "co-constructions".

The fourth and final part of the analysis in section 5 addresses operationally relevant, design-related suggestions for improvement. Section 6 contains concluding remarks.

# 2   Perspectives on standardisation and health care

## 2.1   Standardisation in health care: efficiency and quality

Standardisation is embedded in efforts to improve efficiency and quality in health care (Timmermans & Berg, 1997; Winthereik & Vikkelsø, 2005; Klein, 2003). Given the very significant levels of health care expenditure throughout the Western world, the reasons for concern over efficiency are immediately obvious. The USA spends 14 % of GDP on health care (Light, 2000), the average in the EU is 8.6 % (BBC News, 2000) and in Norway it is 10.3 % of GDP (WHO, 2003). Despite efforts to contain it, expenditure keeps increasing. To illustrate from the Norwegian context, the period 1980 to 1995 saw a 1.2 % inflation-adjusted increase in expenditure per year in somatic hospitals, which increased to 4.8% per year between 1995 and 2000 (SHD, 2000-2001).

Ensuring sufficient quality of treatment and care is another pressing issue for health authorities. In a US Institute of Medicine report from 2000, the Committee on Quality of Health Care in America estimates that medical errors (e.g. errors in administering drugs or planned treatments) are the leading cause of death in the United States (Kohn et al., 2000; IOM, 2001). Similarly, investigations in Norwegian health care indicate that fatal adverse drug events represent a major problem in hospitals, especially in elderly patients with multiple diseases (Ebbesen et al., 2001). It is also suggested that between 5% and 10 % of all hospital admissions are caused by the wrong use of medication (Buajordet et al., 2001; Ebbesen et al., 2001).

Improving both efficiency and quality is an enormous undertaking, especially since the notions of efficiency and quality are often seen as contradictory terms (Law, 2003, p. 10). Yet standardisation is a key element in attempts to improve efficiency and quality in health care. Governmental efforts to achieve standardisation take on many forms. Timmermans and Berg (2003) distinguish between four broad categories of standards: design standards, performance standards, terminological standards and procedural standards. Design standards represent detailed and structural specifications of social and technical systems, ensuring compatibility, logistics and integration. Performance standards represent outcome specifications, identifying the result of an action. An example is the Norwegian initiative in 2003 to establish national quality indicators as an external benchmarking measurement. The purpose was to present to the public a summary of which hospitals could provide the best quality of treatment and care, and to facilitate the growth of market mechanisms in health care.

The third and fourth categories are the most relevant ones for this paper. Terminological standards have had an important role in modern medicine for a long time. For example, the global World Health Organisation (WHO) based ICD[1] (International Classification of Diseases), NANDA[2] (North American Nursing Diagnosis Association) and SNOMED[3] (the Systematised Nomenclature of Medicine. Standardised terminologies have been developed and used to ensure consistency of meaning across time and place, enabling large-scale planning opportunities for local users as well as for national health authorities and international health organisations.

The fourth category of standardisation is derived from the ongoing process of standardising medical work through clinical guidelines, protocols and care plans[4]. The purpose is to establish 'best practices' to "delineat[e] a number of steps to be taken when specified conditions are met" (Timmermans & Berg, 2003, p. 25). Such standards are assumed to increase both quality and predictability, thus "maximis[ing] the likelihood that the same thing is being done to each patient" (Coiera, 2003, p. 146), and also taking account of the cost factor:

> "Over the past three decades, public and private purchasers turned to managed care plans to stimulate greater hospital competition and reduce hospital expenditures and costs" (Devers et al., 2003, pp. 419-420).

## 2.2   Nursing: plans as standardised care

The implementation of electronic clinical nursing plans is an example of the third and fourth type of standards outlined above: terminological standards and the standardisation of medical work. Nursing plans are closely aligned with health authorities' aspirations for quality assurance and cost control. For the nursing profession, however, there is an additional agenda associated with the professionalisation and legitimisation of nursing:

---

[1] http://www.who.int/classifications/icd/en/ (accessed March 20. 2006)

[2] http://www.nanda.org/ (accessed March 20. 2006)

[3] http://www.snomed.org/snomedct/ (accessed March 20. 2006)

[4] We use the terms 'care plan' and 'nursing plan' as equivalents throughout this article.

"Ultimately, the documentation practices reflect the values of the nursing personnel." (Voutilainen et al., 2004, pp. 79-80)

Traditionally, nurses have struggled to achieve status for their profession as independent from rather than subordinate to physicians, and hitherto nurses' documentation has been relatively 'invisible' (Bowker et al, 2001; Star & Strauss, 1999). An effective nursing classification system can therefore be seen as a precondition for the increased professionalisation of nursing.

Care plans are integral to this initiative. Basically, a care plan is an overview of probable nurse-related diagnoses or problems associated with a particular patient group, combined with relevant interventions. It is perfectly aligned with the expectations of increased efficiency and quality outlined above:

"It is expected that nurses obtaining appropriate and accurate information when they need it will improve the chance of making better decisions about patient care." (Lee & Chang, 2004, p. 38).

Similar expectations are echoed in Norwegian policy documents (KITH, 2003a, pp. 10-11; Nurses' Forum for ICT, 2002). The latter argues that:

"An EPR may easily present current guidelines or procedures and then it is possible to document just the deviation (…) this may simplify the documentation and increase the quality of nursing" (Nurses' Forum for ICT, 2002, p. 17).

At the core of the nursing plan is its shared terminology. As with the ICD for physicians, the classification systems embedded in the nursing plan are tailored to nurses' work. Nurses apply this terminology to describe the patients' problem (i.e. nurse diagnoses): they link each problem with one or several interventions, detailing what to do in particular situations.

Some of the best-known systems are NANDA (North American Nursing Diagnosis Association), NIC[5] (Nursing Intervention Classification), NOC[6] (Nursing Outcome

---

[5] http://www.nursing.uiowa.edu/centers/cncce/nic/ (accessed March 20. 2006)

[6] http://www.nursing.uiowa.edu/centers/cncce/noc/ (accessed March 20. 2006)

Classification) and ICNP[7] (International Classification on Nursing Practice) (Hellesø & Ruland, 2001).

In contrast to the ICD, which is more than a hundred years old, classification systems for nurses are a relatively new phenomenon. The first initiative dates back to the early 1970s, when the North American Nursing Diagnosis Association developed NANDA (McCloskey & Bulechek, 1994). Today, further development of NANDA is based on consensus decision-making. Every second year, diagnoses are presented and validated at NANDA conferences. The most recent edition of NANDA, from 2005-2006, contains 167 diagnoses classified into nine domains. Each diagnosis has the following attributes: a label, a definition, defining characteristics and related factors[8].

Both NIC and NOC can be used together with NANDA, as the three systems cover different parts of the nursing process (NANDA applies to problems, NIC to interventions and NOC to outcomes). The NIC taxonomy was developed by the Iowa Intervention Project, which was established in 1989. The first version of the NIC classification was published in 1992, and it is updated every fourth year[9]. The current version was published in 2004 and contains 514 nursing interventions grouped into 30 classes and 7 domains.

Nursing care plans have gained widespread international attention recently, especially with the implementation of electronic patient records (EPRs) in hospitals (Lee, 2005; Lee et al., 2002; Timmons, 2003; Getty et al., 1999; Lee & Chang, 2004). This is because EPRs are recognised as convenient vehicles for formalising nursing work and documentation as well. This trend is evident in Norway (DIPS, 2005; KITH, 2003a; 2003b; Hellesø & Ruland, 2001). However, given the high expectations and extensive initiatives outlined above, the actual *use* of care plans has so far been disappointing. Studies have indicated that "nurses have problems integrating the nursing process and care planning into their daily record-keeping" (Björvell et al., 2002, p. 35). In a survey cited by Sexton et al. (2004, p. 38), "nursing care plans were referred to in handover only 1% of the time and this was probably because care plans were not being updated". One explanation may be that the "nursing process is thought to be time-consuming to document" and its value was questioned (Waters, 1999, p. 80). For instance, some observers have argued that care plans were more significant for the professionalism of nurses than for patient care (Lee & Chang, 2004). In other cases, cultural differences caused difficulties in using a global classification system such as NANDA (Lee et al., 2002).

---

[7] Without going into the matter in depth, we recognise that there are other classification systems for nursing diagnoses and practice as well. These include the CCC (Clinical Care Classification), previously known as the HHCC (Home Health Care Classification) System (http://www.sabacare.com/ (accessed March 20. 2006), the Omaha system and the Patient Care Data Set (Hyun & Park, 2002, p. 100)). The ICNP covers the whole range of diagnoses, interventions and outcomes (Hellesø & Ruland, 2001). A project to establish ICNP was initiated in 1989 by the International Council of Nurses as an effort to unify the existing nursing languages (Hyun & Park, 2002). The ICNP is still a 'young' system, as version 1 was launched in 2005 by the Taiwan International Council of Nurses (http://icn.ch/index.html (accessed March 20. 2006)).

[8] http://www.nanda.org/ (accessed March 20. 2006)

[9] http://www.nursing.uiowa.edu/centers/cncce/nic/ (accessed March 20. 2006)

Due to the infrequent use of care plans, they have not been discussed extensively. Some notable exceptions exist: Bowker et al. (1995) related the NANDA and NIC terminologies to the legitimacy and visibility of the nursing profession. While thoroughly covering these terminologies, they do not describe the actual work of nursing in much detail. Wilson (2002), on the other hand, analysed a case from a UK-based hospital where the nurses rejected a care plan system because it was never associated with nursing. The nurses argued that the system made them prioritise record keeping at the expense of delivery of care.

## 2.3 Standardisation as co-constructive practice

We develop our analytical perspective on standardisation in two steps. First, we discuss the traditional approach to standardisation, which focuses imposing standards top down in a fairly prescriptive manner. The key points here are that the standard is fixed and the users merely adapt to the standard. Secondly, we move from this traditional approach to a *co-constructive* perspective in which standardisation and work practice mutually shape and constitute each other. We thus emphasise standardisation as a socially constructed negotiation *process*.

The traditional approach to global standardisation assigns a very important role to international standardisation bodies as providers of standards. The International Organisation for Standardisation (ISO[10]) based in Geneva, Switzerland, is one of the most important of these bodies, and represents more than 140 countries (EHTEL, 2002). Since NANDA and ICNP have a global scope, they have both asserted compliance with ISO 18104.

---

[10] http://www.iso.org/iso/en/ISOOnline.frontpage (accessed March 20. 2006)

Another international standardisation body important in healthcare is the HL7[11] accredited by the American National Standards Institute[12]. HL7 is today the largest health information standards developer in the world. It focuses on the electronic interchange of clinical, financial and administrative information among independent healthcare-oriented information systems (Tsiknakis et al., 2002, p. 11).

For the European healthcare sector, the CEN/TC251 is a major standardisation body. It is responsible for organising, coordinating, and monitoring the development of standards in health care (van Bemmel & Musen, 1997, p. 515). In Norway, standardisation in health care is coordinated by the Norwegian Centre for Informatics in Health and Social Care (KITH).

A striking feature of these organisations is that their scope on standardisation is extended. From dealing with technical standards and terminologies, they are now increasingly "interlocking with and being reinforced by the drive toward evidence-based medicine" (Timmermans & Berg, 2003, p. 7). This implies that processes, work practices and guidelines are of increasing concern. An illustration is the ISO standard IWA-1 (2005), which aims at:

> "provid[ing] additional guidance for any health service organisation involved in the management, delivery, or administration of health service products or services, including training and/or research, in the life continuum process for human beings, regardless of type, size and the product or service provided" (ISO IWA, 2005).

Still, the common strategy for both international and national standardisation agencies is to develop standardisation far away from local work practice. Sometimes, local work practice is even defined as the real obstacle to standardisation. For example, the former chairman of CEN/ TC 251, De Moore (1993) asserts firmly that it is important to eliminate standards evolving from local contexts:

> "to make sure that unsuitable circumstances (e.g. proliferation of incomplete solutions) are not allowed to take root…[so] standardisation must be started as soon as possible in order to set the development in the right track" (De Moore, 1993, p. 4).

---

[11] Health Level Seven at http://www.hl7.org/ (accessed March 20. 2006)

[12] American National Standards Institute at http://www.ansi.org/ (accessed March 20. 2006)

However, a major flaw in this position is that it downplays to the level of non-existence the challenges of implementation, i.e. the *process* of standardisation (Akrich, 1992). Empirical studies demonstrate vividly how political negotiations influence standardisation processes (Bowker & Star, 1999; Lachmund, 1999; Hanseth & Monteiro, 1997). Bowker and Star (1999, pp. 120-121) use the example of the issue of still births in the 1920s: Catholic countries fought to recognise the embryo as a living being, statistically equivalent to an infant, while Protestant countries were far less likely to do so. Similarly, Hanseth and Monteiro (1997) describe how the emergence of standards for exchanging laboratory results between laboratories and general practitioners saw different arguments framed as trade-offs between different technical costs and benefits, while the real issue at stake was a race between different actors, promoting technologies which seemed most beneficial for them.

Through their work on clinical protocols, Timmermans and Berg (1997) argue similarly that while standards attempt to change and replace current practices, they also need to incorporate and extend those routines. The standard is expected to function in a work practice consisting of existing interests, relations and infrastructures.

Timmermans and Berg (1997) also point out that users are anything but mindless slaves to standards. Rather, minor and not so minor deviations are practiced routinely. They describe other tinkering strategies to make the protocol work, such as searching for the right protocol for their patient, introducing deviations and adaptations, and even circumventing the protocol. At times the users go beyond the boundaries of the protocols, making ad hoc decisions and even repairing the deviations of others. However, an important point is that such tinkering with the protocol is not a failing, but a *prerequisite* for the protocol to function: it allows leeway to adjust the protocol to unforeseen events (ibid, p. 293).

To sum up, the design and use of a standard are *co-constructed*. In this way, the global standard both shapes and is shaped by local work practice. In the words of Timmermans and Berg (1997, p. 297), standardised work always involves 'local universalities'. Our contribution in this regard is that we combine this theoretical insight with an in-depth empirical study, demonstrating how standardisation unfolds in practice.

# 3 Method

## 3.1 Research setting

The Department of Special Psychiatry (SPA) is located in the countryside outside Tromsø, some 5 kilometres away from the rest of the University Hospital in Northern Norway (UNN). It is the only institution in the health region which accepts involuntary admissions of patients suffering from psychiatric disorders. The department's area of expertise encompasses psychogeriatrics, drug addiction associated with serious psychiatric problems, and aggressive behavioural disturbances, including patients with sentences imposing psychiatric therapy. Approximately 350 people work at the institution, which admitted 155 patients in 2005.

Our study was carried out in the psychogeriatric ward at the Department of Special Psychiatry. Patients in this ward are aged 65 years or more, and suffer from dementia, senility or anxiety. The ward has 15 rooms, and treats 95 patients a year with an average length of stay of 6-8 weeks. Some 45 people work permanently here, including nurses, unskilled workers and substitutes[13], social workers, occupational therapists and physiotherapists. In addition, three physicians and one psychologist pay regular visits. The staff turnover in the ward is high, with up to 5 new unskilled workers starting each month.

In the day room, one often finds nurses talking quietly to the patients, in a calming manner. However, this may change as one patient suddenly starts to yell and shout, unable to control his anxiety or aggression. Then additional nurses are quickly called for and a set of predefined measures is put into action.

Due to the somatic and psychiatric complexity of the patients' conditions, the ward relies on an interdisciplinary approach to treatment and care. Nursing observations are particularly important, as one of the physicians explained:

> "In this ward, medical treatment has little effect on the patients. Therefore, environmental therapy becomes especially important (…). Several of our patients come from closed units and have a history of smashed doors and walls. After a couple of days in here they are meek as a lamb." (Physician)

---

[13] Unskilled nurses and substitutes fill the same role as qualified nurses and are referred to as nurses in this paper.

## 3.2   Research method

Adhering to an interpretative research approach (Klein & Myers, 1999; Walsham, 1995), our main aim was to understand the standardisation of work as it unfolds in the practice of everyday nursing. Data collection methods consisted of i) semi-structured interviews, ii) participant observations and informal discussions, iii) document analysis, and iv) participation in internal project meetings.

Fifteen interviews were carried out between May and December 2005, at ten of which two of the authors were present. On average the interviews lasted 1-1.5 hours. They were taped and subsequently transcribed.

| | Field trips | 1st visit | 2nd visit | 3rd visit | Total |
| Position | | May | June | August | |
|---|---|---|---|---|---|
| Norwegian Nurses Association | | 1 | | | 1 |
| Project manager (hospital) | | | 1 | | 1 |
| Nurses | | 1 | 3 | 4 | 8 |
| Physician | | | 1 | | 1 |
| Psychologist | | | | 1 | 1 |
| Project group nurses (local) | | 1 | 1 | 1 | 3 |

**Figure 1: Categorisation of the 15 interviews involved in our study**

In total, 80 hours of observation were conducted, mainly in the duty room during reporting, but also during other activities such as nurse handovers, interdisciplinary cardex and treatment meetings. Handwritten field notes were written up as soon as possible after each observation session. While observing, we attempted to cover a range of actors and interactions. For instance, in the observation of work activities and discussions, we looked for potentially different interpretations of the same phenomenon.

The third and fourth methods of data collection were document analysis and participation in internal EPR-project meetings. This included both collecting and reading relevant documents about the project itself (specifications, news letters, training material) as well as the nursing documentation (reports, plans and cardexes). During the second and third visit to the ward, we also attended four internal EPR project meetings where we were increasingly able to provide feedback on our findings.

The overall process of collecting and analysing data was open-ended and iterative, with the earlier stages being more explorative than later ones. Empirically and analytically, all three authors have an extensive history of involvement in the health care domain, including a shared interest in the design and use of EPRs. The first author has studied the implementation of EPRs at UNN for several years. The second author has a long history of involvement in national and international projects dealing with health information systems. The third author has been following the implementation of electronic nursing documentation at three Norwegian hospitals in addition to UNN.

Our analytical categories emerged gradually from internal discussions, reading of field notes and external presentations. However, first-order conceptualisation (van Maanen, 2002) started at the field site. When possible, we reflected on our observations and discussed potential issues to pursue further. At the end of each day, we discussed our observations and made plans for the following day. Between each field trip, notes from our individual observations, transcribed interviews and collected documents were shared and discussed. An important product of this work was a document of second-level issues and concepts (van Maanen, 2002), which was also used in discussions with the second author.

During our first field visit we spent a significant amount of time engaged in informal discussions with key actors in the project, partly to gain legitimacy, and partly to inform the issues of our study (Klein & Myers, 1999). Plans for our field study were made in cooperation with the head of the department. Having generated general insight into the project during our second field visit, we directed our attention towards the psychogeriatric ward. At this stage, theories on standardisation (Timmermans & Berg, 2003) had been identified as a major theme for our study, thus guiding our data collection strategy (see Principle of abstraction and generalisations in Klein & Myers, 1999, p. 72).

The combination of observation and interviews was particularly useful both to validate our observations and to provide access to data that was not otherwise readily available.

We validated our interpretations by presenting preliminary results at several seminars. Firstly, we presented our findings to the staff using the EPR in the Department of Special Psychiatry. Secondly, we presented and discussed our findings on two occasions with research colleagues at the Norwegian EHR Research Centre (NSEP). Finally our work was presented to the full executive board of the vendor of the EPR, which we will refer to as 'HealthSys'.

# 4  Case: standardisation of nurse work

## 4.1  Motivation and start-up

During 2004 and 2005, the University Hospital of Northern Norway (UNN) was the site of a large-scale EPR implementation project. The aim was to establish a common EPR infrastructure which cut across departmental (clinic and laboratories) and professional (physicians, secretaries and nurses) boundaries.

HealthSys is a major Norwegian-based vendor of health-based information systems, currently serving about a third of the Norwegian EPR market. In addition to the EPR, HealthSys offers laboratory systems, patient administration systems and radiology systems. Together, these systems are promoted as parts of an all-encompassing hospital system. The project at UNN started in 2003 with the decision to acquire all of the HealthSys modules. Figure 2 illustrates local initiatives at the Department of Special Psychiatry in the context of hospital level initiatives.

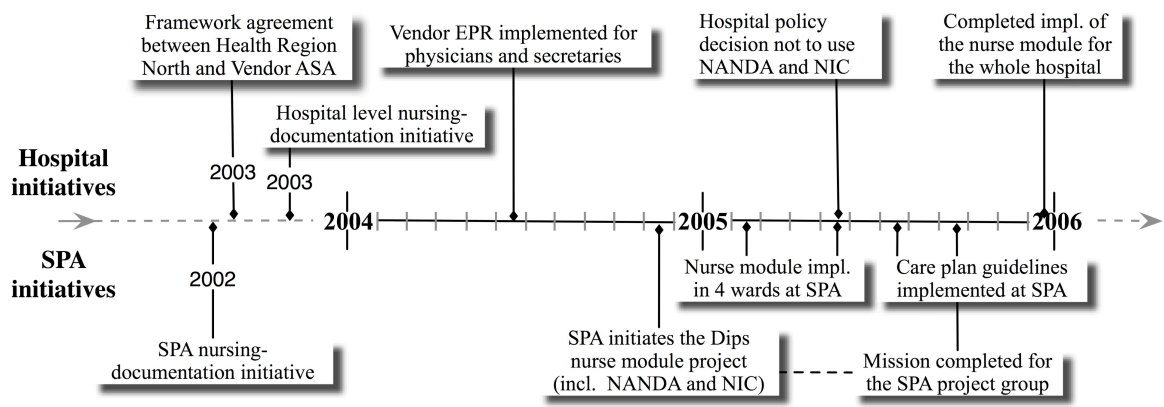

**Figure 2: Timeline illustrating EPR initiatives at the Department of Special Psychiatry (below the timeline) in the context of hospital level initiatives (above the timeline)**

A major goal in the project has been to replace the existing paper-based nursing documentation. The HealthSys EPR contains a nursing module which has been developed to support the nurses' daily reports (several per day) and a structured nursing plan which supports planning and overview. The electronic nursing module was implemented in the Department of Special Psychiatry in April 2005 (an example of the interface is provided in Figure 3).

The Department of Special Psychiatry was highly motivated to implement the nursing module in its four wards. The departments' keen interest was associated with increased political attention towards improved quality in the psychiatric sector. At the same time, the Norwegian Nurses Association was interested in promoting the nursing profession in the health sector. The nursing plan was seen as a means towards achieving this goal (NSFID, 2004; cf. section 2.2). National interests and rhetoric were thus translated into local demand for improved documentation practice:

"We must concentrate on documenting what is important and exclude [details such as] whether someone has eaten four slices of bread with jam or whether the husband brought five roses when he paid a visit" (project group nurse 1)

Some of the nurses even suggested that the nursing plan might contribute to improved efficiency and a better overview of the process of planning. In that sense, the former content of the reports, such as diagnoses, interventions and other repetitive patient-related information, could be transferred from the reports to the nursing plan:

"In fact, if you are involved in planned care, you should hardly have to write daily reports at all as everything should be in the nursing plan. For instance, if the plan states that the patient needs help related to feeding and anxiety (…) and we adhere to it each time, we need not reiterate this in every report" (project group nurse 2)

In the spring of 2004, the Department of Special Psychiatry conducted a workshop on electronic nursing documentation. The vendor, HealthSys, also participated. In November 2004, the department established a project with the aim of implementing the EPR nursing module. Two nurses and one secretary were recruited internally to run the project. They spent two days a week preparing for the implementation of electronic nursing documentation in the department's four wards. This included training users who lacked basic computer skills, regularly coordinating activities between the local project and the central EPR project at UNN, and reading reports from and visiting other hospitals engaged in similar projects. They also developed a help system for basic nursing procedures in the new EPR.

In sum, this contributed to a relatively smooth start-up process of the system in February 2005, both in the psychogeriatric ward and in the three other wards in the department.

## 4.2   The nursing module in the new EPR

For each patient there is only one nursing plan. Basically, the nursing module is divided into two very different parts. The first part is the report section where users write reports on a patient several (usually three) times a day. Although there is some structure in this section (see Figure 3), the users have the flexibility of writing free text, i.e. constructing a narrative of the patients' problems. The second part is the nursing plan section consisting of international codes, identifying diagnosis and related interventions for a patient. In spite of the difference between the report and the nursing plan, they are interconnected and mutually dependent. Each time a report is written or read in the upper part of the screen, the patients' current nursing plan is presented in the lower part.

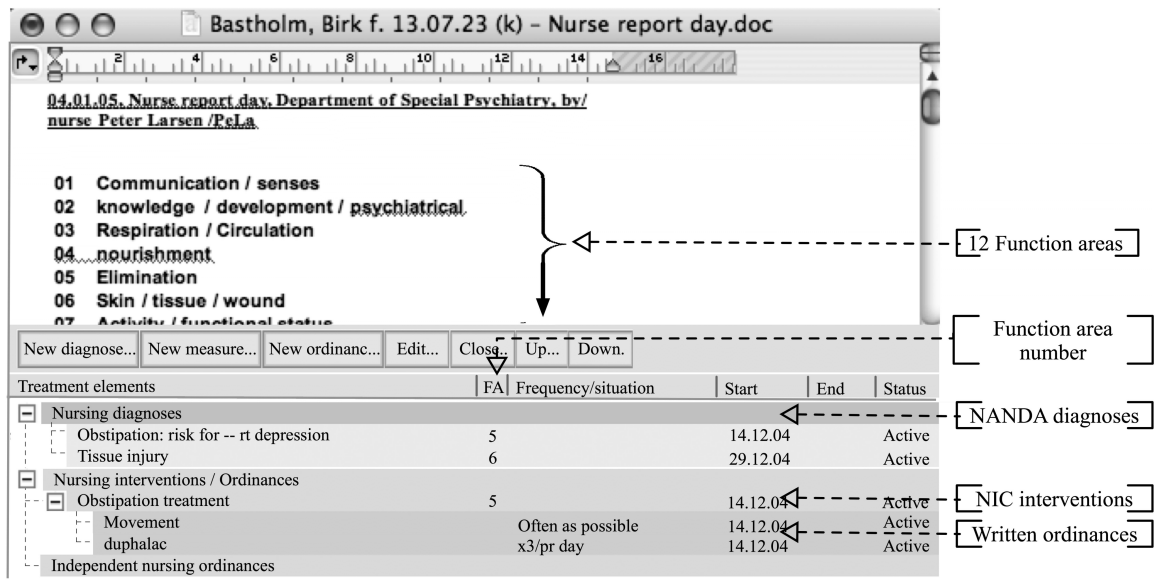

**Figure 3: The interface of the nursing module. Each time a report is written, the current care plan is presented in the lower part of the screen. The function area number represents a possible connection between the report and the plan.**

The content of the report is structured according to the 12 function areas (see Figure 4).

| 01 | Communication / senses |
| 02 | Knowledge / development / psychiatric |
| 03 | Respiration / Circulation |
| 04 | Nourishment |
| 05 | Elimination |
| 06 | Skin / tissue / wound |
| 07 | Activity / functional status |
| 08 | Pain/ sleep / rest / well-being |
| 09 | Sexuality / reproduction |
| 10 | Social / planning of discharge |
| 11 | Spiritual / cultural / lifestyle |
| 12 | Other things / tasks delegated from physicians and observations |

**Figure 4: The 12 function areas in the report, which also is used to organise NANDA diagnoses and NIC interventions in the care plan.**

The nursing plan is based on the NANDA and NIC classification systems. One NANDA diagnosis may spawn one or several NIC interventions. For each NIC intervention there may be several instructions (direct actions). The instructions are written in plain text extensions in the plan.

NANDA and NIC are structured into 12 function areas in the report. This is not a part of the international NANDA and NIC classification schemes; it has been introduced by the vendor, HealthSys, to make it easier to find specific diagnoses and interventions in a given function area. The function area for a given NANDA/ NIC code is also shown in the plan, indicated by a number. This makes it easier to write a report and indicates the categories in the report to be filled in, based on the function areas in the plan. More specifically, the user writing the report is expected to use the plan with its diagnosis, interventions and instructions as a basis for the reports. Only deviations from the plan are expected to be documented in the report, thus keeping the content of the report to a minimum.

> "The goal is to write as little as possible in the report and to write in relationship to what is in the nursing plan and describe deviation from it" (Project group nurse 2)

A NANDA diagnosis and a NIC intervention may fit within several function areas, emphasising the challenge of finding the links between the two classification systems. There is no formal connection between diagnoses and interventions, because one diagnosis may require several interventions and one intervention may cover several diagnoses.

## 4.3   The use of the nursing module in the psychogeriatric ward

The users emphasise two key outcomes of the electronic nursing documentation project in the psychogeriatric ward. Firstly, the EPR nursing module implementation is generally perceived as a success as it provides a clear overview:

> "People attending the meetings have already read the reports, nursing plans and everything. So now we focus on the core of the case (…) and don't have to read everything aloud during the meetings" (Nurse 1).

An experienced nurse on regular night duty, covering all the four wards in the department, elaborated:

>  "Now, when I come to the psychogeriatric ward (…), I just open the nursing plan and see the diagnoses instantly. Since the plan contains standardised codes, I get a quick overview of the patients' troublesome areas, thus informing me of what to expect" (Project group nurse 3)

From this rather positive outcome we move on to the second consequence of the implementation. In spite of a well-planned project, the project members had to cope with an unexpected situation: the users realised that adding new diagnoses and interventions required intensive mouse-clicking through various windows, dialogue boxes and menus in the application. In short, they felt that the user interface was not user-friendly (enough):

> "You have to [actively] select a function area, which should have appeared automatically … then you have to respond to 'Do you want to save?' repeatedly … [also] there are poor search possibilities when removing interventions" (Nurse 2)

Furthermore, even though the NANDA diagnoses and the NIC interventions represent relatively wide categories, additional work was required to find the right category. The users had to spend significant amounts of time searching for diagnoses and interventions.

Another difficulty was that the broadness of the categories made the codes useless as standalone codes:

> "By themselves, the codes are completely open and many of them say absolutely nothing" (Nurse 3)

## 4.4   The EPR nursing module and the broader context

While the psychogeriatric ward was positive about using the EPR nursing plan, the other three wards in the Department of Special Psychiatry were more reluctant. We believe that the reluctance may be understood in two ways. Firstly, the *turnover* frequency of the patients in the other wards was not as high as that in the psychogeriatric ward. Consequently, these patients and their needs were already known, and there was less need for communication and overview. Moreover, the plan in one of the wards (the security ward) had a different role to that of the nursing plan in the EPR. Their plans, straightforward A4 paper sheets, were 'negotiated contracts' between the staff and the psychiatric patients.

Secondly, reluctance to use the EPR in the other wards must also be understood in terms of how the classification systems NIC and NANDA were assumed to imply fragmentation of the nurses' work in general. Classification systems for nurses were considered to be a threat to the traditional holistic way in which nurses provided care to patients. This view was not confined to wards in the Department of Special Psychiatry; it mirrored a concern in many of the other departments participating in implementing the nursing module. At a meeting for the hospitals' head nurses in May 2005, one of them asked rhetorically:

"Will there be two languages now, one for the clinic and one for research and statistics [based on NIC and NANDA]?" (Head nurse 1, from another department)

As a result, the group of head nurses decided not to proceed with NIC and NANDA at that time. This decision, however, did not influence the project at the Department of Special Psychiatry as they already had been using the EPR nursing module for several months. The lack of enthusiasm for nursing plans in the rest of the hospital meant that when the implementation of the nursing module in December 2005 was completed, it was only the Department of Special Psychiatry (or more precisely, the psychogeriatric ward), which had gained in-depth experience of the new system.

# 5   Analysis

The purpose of our analysis is to map and discuss how structurally imposed standardisation efforts mesh with the everyday practice of health care delivery. Our point of departure (see section 2) is that the standards to be imposed had the status of intentions. They were embedded or institutionalised into work routines through a process of *transformation* – in part intended, in part non-intended – of both the standards and configurations of work. In this sense, standardisation needs to be recognised as *co-constructed practice*.

## 5.1   The invisible work of fitting categories

The core idea of a plan-based approach to nursing at UNN is to work out a list of pairs of NANDA/NIC for every patient. In other words, the plan consists of a number of pairs where each pair consists of one nurse diagnosis (coded in NANDA) tied to one intervention (coded in NIC). Despite this conceptual simplicity, a lot of non-obvious work is involved in establishing each of the NANDA/ NIC pairs. This corresponds closely to what feminists termed *invisible* work, and which subsequently has been identified in numerous settings and forms by IS scholars (Schmidt & Bannon, 1992; Star, 1991). For instance, Bowers (1994) points out the essential, yet 'invisible', element of maintenance and constant support throughout the implementation process.

Establishing the NANDA/ NIC pair involves a substantial amount of relatively time-consuming searching. Typically, one might start off by attempting to determine diagnosis code. NANDA, however, contains 167 distinct diagnosis codes, which are difficult to remember. There are two ways to search for NANDA codes at UNN. The user can search directly by entering the first letter of the word of the diagnosis. Alternatively, the user may delimit the returned values by selecting a function area (see Figure 5, which illustrates the selection of function area 3). By choosing a function area, the user retrieves approximately 13 of NANDA's 167 diagnoses. It is then relatively easy to browse through all the diagnoses in this category.

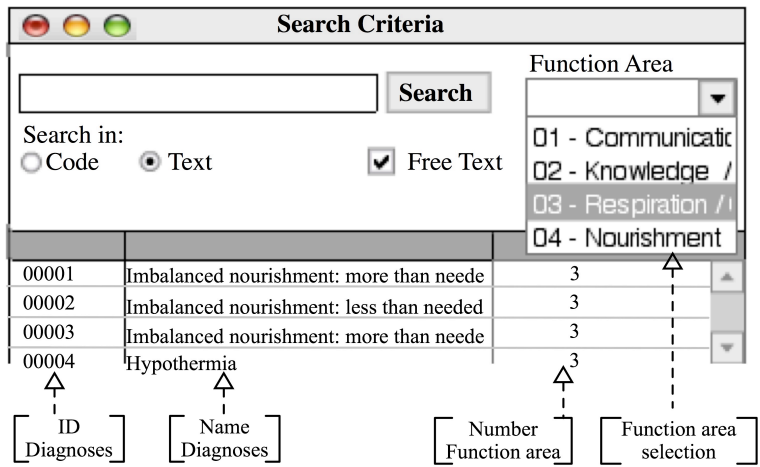

**Figure 5: Interface for searching among NANDA diagnosis codes. To search, the user can either use free text or select one of 12 function areas. The figure shows the latter search method.**

The search for NIC interventions is accomplished in a similar way, but there are 514 NIC interventions. The training manual in the department suggests that even if a function area is selected, there are still too many interventions to browse through, and hence encourages the use of a search word before the search button is used. Still, the exact match is sometimes difficult to obtain:

> "Sometimes you don't find the interventions you need and end up having to take what is closest. You also have to say things in other words. In addition, you have to say how often an intervention is going to occur. The biggest difficulty is that you cannot write freely". (Nurse 4)

There is no formal relationship between diagnoses and interventions, because one diagnosis may require several interventions and one intervention may cover several diagnoses. The only link is through the associated function area. As a NANDA diagnosis and a NIC intervention may belong to, and cover, several function areas, the challenge of identifying the links between the two classification systems tends to be time-consuming.

Both NANDA and NIC are constructed as general-purpose classification schemes, i.e. are intended to cover all types of (western) hospitals. As the ward we studied is highly specialised, this implies that only a subset of the total NANDA and NIC codes are relevant; the codes used are clustered around only a small proportion of the 167/514 which are available. Specifically, function area 02 (see Figure 4) addressing 'Knowledge / development/ psychiatric' is favoured in a majority of cases "as a general rule" (Nurse 3). Moreover, the relatively few codes within function area 02 are too crude to capture the variations in practice in the ward.

The relatively open categories within function area 02 are not precise enough to inform the subsequent actions which are planned. In response, the nurses actively refine the broad categories by adding '--' and a subsequent amendment. Figure 6 illustrates how open categories are broken down and specialised by filling in '--' and free text. In the plan these amendments appear directly after the NANDA diagnoses in the plan, marked by '--'. In figure 6, six of the eleven diagnoses have an amendment.

**Figure 6: The figure illustrates how open categories are broken down and specialised. The double hyphen '- -' is used to separate free text information (nurses' elaboration) from standard text (NIC interventions).**

One nurse explains the need to add details to the NANDA diagnosis 'Risk for violence against others' (see Figure 6):

> "Look here, this is the diagnosis 'Risk for violence against others', but we have to add 'verbal threats', 'threatening behaviour when we activate restrictions for him'. We have to add these things to understand the patient" (Nurse 4)

The practice of elaborating the given categories through specialisation illustrates a general dilemma concerning the trade-offs in calibrating the level of *granularity* in schemes of categories and standards: a crude level of granularity (i.e. open categories) imply that relatively little work is required when writing, but a corresponding amount of work for the reader is required and *vice versa*. Building empirically on the ICD, Bowker and Star (1999) elegantly explain a similar trade-off between general practitioners (writers) and health policy institutions (readers).

## 5.2   Transforming categories

The classification schemes of NANDA and NIC are not merely tinkered with or adjusted marginally as illustrated above. They are transformed and reconfigured actively by the nurses through their gradual institutionalisation. This goes well beyond reactive 'adaptation' and indicates what Berg and Timmermans (2000, p. 45) accurately identify as the *constitutive* element of the users. The users of classification schemes are not meek subjects of an imposed standard; they participate in altering – ultimately transforming – that very standard. Without this transformation, the standard would not work.

A practical and real concern, especially for the more elaborate plans, is to maintain a clear sense of which diagnoses are linked to which interventions. In the current system, the only way to make these connections was via the function areas. Figure 7 illustrates how the NANDA diagnosis 'Anxiety' (circled in the yellow part) is linked to three different interventions (circled in the blue parts). Given this, a key concern was to manipulate the sequence of the diagnosis and interventions to ease readability in general and communicate in a more nuanced way about degrees of urgency in particular. The patient whose plan is depicted in figure 7 suffered from numerous conditions, but one of the most important was neglect[14] on her left side caused by a stroke. She therefore became extremely anxious, which resulted in frequent shouting and yelling. As the nurse working out the plan said while pointing to the NANDA code 'Anxiety':

> "To reduce anxiety is *the* most important intervention to avoid the shouting and
> yelling (…) the problem is that the intervention related to this diagnosis appears so far

---

[14] After a stroke, some people suffer from a syndrome called 'neglect' People with neglect may not appear to process sensory information from the left side of their body. http://ww2.heartandstroke.ca/Page.asp?PageID=33&ArticleID=2570&Src=stroke&From=Sub Category

down in the plan that it is difficult to see the relationship between the diagnosis and the intervention" (Nurse 3)

To emphasise and communicate this to subsequent readers of the plan, she moved this intervention (broken line in Figure 7) to the top of the list of NIC interventions to signal utmost urgency.

| Treatment elements | FA | Frequency/situation | Start | End | Status |
|---|---|---|---|---|---|
| Nursing diagnoses | | | | | |
|   Anxiety -- rt confusion | ② | | 09.08.05 | | Active |
|   Impaired mucous membrane | 4 | | 30.08.05 | | Active |
|   Insufficient sleep | 8 | | 10.08.05 | | Active |
| Nursing interventions / Ordinances | | | | | |
|   Reducing anxiety -- Objective: security, patient trust | ② | | 09.08.05 | | Active |
|     Wake up before breakfast | | Always | 30.08.05 | | Active |
|   Encourage sleep | 8 | | 09.08.05 | | |
|     Make sure the patient get enough sleep | | | 09.08.05 | | Active |
|     Consider medication | | Together with physician | 09.08.05 | | Active |
|     Record sleeping pattern | | Make list, record in report | 09.08.05 | | Active |
|     Help patient maintain diurnal rythm | | | 09.08.05 | | Active |
|   Sense of reality | ② | | 09.08.05 | | |
|     Clear messages about what to be done during the day | | Written, Oral | 09.08.05 | | Active |
|   Improve feeling of security -- introduce yourself, tell when you are about to finish your watch, offer contact | ② | | 23.08.05 | | |
|   Heal wound -- No denture lower jaw; objective: prevent wound in the gums | 6 | | 30.08.05 | | ------- |
|   Activity-therapy -- follow week-schedule | 7 | | 23.08.05 | | |
| Independent nursing ordinances | | | | | |

*Anxiety is moved to the top of the list*

**Figure 7: The plan for a patient suffering from numerous conditions. Notice how the intervention 'Reducing Anxiety' has been moved to the top of the list to indicate its increased importance.**

Another way in which the content of and practices around plans were transformed was the way *redundancy* was (re-)introduced to make them more robust. As previous scholars have noted (Perrow, 1984; Ellingsen & Monteiro, 2003), certain instances of redundancy of information, despite contradicting fundamental principles in traditional IS and database design (Bass et al., 2003), fill productive and practical roles in ongoing work.

For instance, although certain information was already contained in the plans, sometimes the daily report repeated the content of the plan. Consider this extract from the daily report which Nurse 3 wrote for a diabetic patient:

"Be aware of the restrictions concerning fruits, cakes, etc. The patient is not capable of regulating the amount of these things. See the nursing plan" (extract from the report)

This information was already captured in the plan, so why repeat it here? Nurse 3 explained:

> "Sometimes things are registered *twice*, that is, what is in the report you may also find in the nursing plan. This has to do with experience… I know that the report is read aloud at the change of shift meeting while the nursing plan is not" (Nurse 3)

In response to an inability to decide uniquely how to classify interventions, a common strategy is to *duplicate* the information by entering it in both possible places, but slightly rephrased to 'cover up' the duplication. Consider the patient with neglect on her left side. The nurse was not quite sure where to put the instructions 'talk to' and 'inform'. She finally decided to place them under the NIC intervention 'reducing anxiety'. But after further reflection, she decided that this instruction might equally well be placed in the NIC intervention category 'neglect – left side'. Therefore, she rephrased the instructions 'talk to' and 'inform' into 'explain what is going on' and added it to the 'neglect – left side' category as well. She admitted that this meant that similar instructions were entered in several categories, but as she said: "It has to be like this in order to be visible in both places" (see Figure 8).

| Intervention: Reducing anxiety |
|---|
| Instruction: **Talk to** |
| Instruction: **Inform** |
| Intervention: neglect – left side |
| Instruction: **explain what is going on** |

**Figure 8: An example of how similar instructions are entered in two different NIC-intervention categories to make them visible in both places.**

A common (but often downplayed) feature of health work is the constant and considerable element of insecurity about what to do next (see e.g. Berg, 1997). Rather than following clear plans, care delivery frequently takes the form of stepwise explorations with high degrees of uncertainty. Consider what the nurse says, having selected a nursing plan for a patient.

> "This patient was so crazy when we admitted her that we did not know what to do. Look what they have done here … it was cunningly done … they have put a question mark ("?") behind the diagnosis because they did not know whether she suffered from hallucinations or not" (Nurse 2)

This signalled that the patients' problem had not yet been defined. In the field for NIC interventions, this was followed up by suggestions. For instance, for the NIC intervention

'active listening', the amendment "with regard to development of dementia, delusion or confusion" was added. This encouraged staff to observe the patient closely. Another strategy was to suggest different options, encouraging the staff to try them in turn:

> "Try out these things to make the patient eat: type of food, specific locations, ask the patient each time what he wants, provide specific remedies to keep the spoon, black briquette under the plate, be present, etc." (Project group nurse 3)

In this way, the staff got to know the patient and to define the problem. After a while the plan would be tightened up and become more precise. In sum, the extent and character of the transformation of the apparently ready-made categories in NANDA and NIC demonstrate their essential, or indeed, constitutive, role.

## 5.3   Relocating disorder

Berg and Timmermans (2000) highlight how the ordering effects sought through processes of standardisation simultaneously produce disordering effects. They argue that "[T]he order and its disorder (…) are engaged in a spiralling relationship—they need and embody each other" (ibid, p. 37). What they suggest is that an information system with an apparently clear purpose may abruptly become something different, serving completely different purposes. Moreover, the system may induce surprising consequences, as the order which the system creates for some aspects creates a corresponding disorder for others. In a similar way, Law and Singleton (2005) argue that objects (information systems) may inherently constitute several realities, and may sometimes be "complex, multiple and (in some cases) mutually exclusive" (ibid, p. 342). We provide two examples of this. Firstly, we illustrate how the standardisation of nursing plans unintentionally subverted the possibilities for interdisciplinary cooperation, i.e. how benefits for nurses simultaneously produced disadvantages for the psychologists and physicians. Secondly, we indicate how the nursing plan abruptly invoked another rationality regarding the purpose of the nursing plan, namely as a resource management tool.

Earlier, we have pointed out how the psychogeriatric ward relied on interdisciplinary work between the nurses on one hand and the physicians and psychologists on the other. The narrative contained in the old reports had been the glue in this collaboration:

> "Several of the nurses sum up in their own words after we have had a treatment meeting [for a patient] (…) they write good and extensive notes, especially when something extraordinary has happened (…) Therefore, when I write my own report I often refer to the report written by the nurse" (psychologist)

Furthermore, in the old paper-based version of the reports, other professionals sometimes added amendments to the reports originally written by one of the nursing staff, thereby making the report more complete. An example from one of the paper-based reports is when a physiotherapist expanded on the comment provided by the nurse who had written that the patient had exercised with the physiotherapist, but soon got tired. The handwritten amendment was inserted just below the nurse's report:

> "The patient followed the instructions more poorly than yesterday, but managed to get up and sit down satisfactorily. He walked a round in the walkway. There was no apparent pain in the thighs and knees" (physiotherapist)

As opposed to the reports, the nursing plan is a distinct tool for the nursing staff, which excludes the participation of physicians and psychologists. Consider how the nursing plan was targeting purely nursing work:

> "Previously, we have been very concerned about mediating what the physician has prescribed, the results of tests, diagnoses, etc, but nothing about how to approach an anxious patient (…) Alternatively, if we make a good nursing plan, we will see the patients' problem from the perspective of the nursing staff" (Project group nurse 1).

The physicians shared the same understanding. One of them commented:

> "In the same way as the nurses don't involve themselves in what kind of medication is given (except for antidepressants and antipsychotic medication), the nursing plan is primarily used by the nurses" (physician)

As the plan failed to support interdisciplinary work, it may also limit the communication between the nursing staff and the patients, which was an important feature of the plans in the security ward (see case description). In this ward, a nursing plan functioned as a contract between the personnel and a patient. Along similar lines, a head nurse from one of the somatic departments reported at the head nurse meeting:

> "We produce documentation together with the patients, and we translate between ourselves and the patients (…) but the patients haven't got a language suited to classification schemes. The question then becomes one of how to deal with this in the future" (Head nurse 2, from another department)

We have elaborated on how the process of creating order for nursing work (the nursing plan) has created disorder for interdisciplinary work (through the reports). Following Berg and Timmermans (2000, p. 36), we argue that the nursing plan does not only perform its own order, it also always contains it. Extensive use of the nursing plan emphasises its role in professionalising nurses' work, but at the same time it undermines the use of reports and thus interdisciplinary work. Conversely, low use of the plan requires extensive use of the reports. Actually, we are facing two mutually exclusive realities of the nursing plan (Law & Singleton, 2005, pp. 342-343).

Along similar lines, an information system may appeal to a new reality, and become something completely different - in this case, the nursing plan turned into a resource management tool. Resource management in the psychogeriatric ward was a complex issue, depending on the current condition of the patient, the legal clauses in effect, the going-out status and follow-up. 'Going-out status' indicates whether a health worker needs to accompany the patient outside the ward or not. 'Follow-up' indicates what kind of attention a patient might need, and how often. Having a good overview of such issues was extremely important, as "suicidal patients can never go out alone, but must always be accompanied by one of the health personnel" (Nurse 5). The rhetoric around the plan was modified to include resource management as well,

> "The ideal situation would be to document going-out status and follow-up in the nursing plan; then we could have an overview of the resources needed and how they developed" (Project group nurse 2)

The users themselves had a key role in the transformation process of the plan. Even if the important factors, going-out status and follow-up, were not explicitly part of the plan, the staff used them implicitly to obtain an overview of the resources needed:

> "By reading this plan, I can see that this patient will require a lot of time and resources" (Nurse 1)

Also in the maintenance of the nursing plan, it became increasingly important to include the resources needed. For instance, when a nurse was updating the nursing plan, one of the project leaders passed by and reminded her to include the staff resources needed:

> "You must include that this patient needs one-to-one follow-up (…) we have to be precise about which resources are needed in order to succeed with the nursing plan" (project group nurse 2)

Although it had been intended primarily as a vehicle for tracking the ongoing delivery of nursing care, the nursing plan implementation process became increasingly entangled with managerial concerns for resource management and control. The use of clinical information was thus lifted out of its primary context in order to be used for completely different purposes.

## 5.4  From 'Implications for design' to interventions

Interpretative studies of the use of information systems, typically geared towards in *situ* descriptions of work practices and user perceptions, are potentially a rich resource for determining the requirements for system design. There is an old and ongoing debate regarding the exact nature of this relationship (Hughes et al., 1992; Plowman et al., 1995). In a recent and poignant instance, Dourish (2006) is concerned about the strong tendency for interpretative studies in general and ethnographic studies  (including ours) in particular to be reduced to a mere "toolbox of methods for extracting data from settings", so  "aligned with the requirements-gathering phase of a traditional development model" (ibid., p. 543).

We agree with Dourish's general concern about making ethnographic research more relevant for design of information systems. However, an implicit assumption in his formulation of the 'problem of implications for design' is that he is only concerned with the requirements themselves (i.e. their content and functions). In this sense, the 'problem of implications for design' has a strong bias towards the local and singular work setting. This downplays how the researchers negotiate their results in distinct *arenas* with different stakeholders.

Inspired by insights from studies in the fields of science and technology, we move beyond localised, artefact-centric 'implications for design' to network-based *interventions* (Braa et al., 2004). Rather than handing over a context-free set of requirements derived from our study ('a bullet list' (Dourish, 2006, p. 549)), we must consider the requirements as just one element that needs to be negotiated with the stakeholders in distinct arenas. Our results were tailored to the different needs and expectations of stakeholders; in other words, we varied the form, granularity and perspective significantly across arenas. This variety mirrored power relations and different expectations towards us as researchers. This "provide[d] the opportunities to build the relationships to forge alliances across potentially disparate interests" (Balka, 2005, p. 12), thus making our research 'practically relevant' (Bal & Mastboom, 2005, p. 7).

We identified and subsequently intervened in five different arenas. These were associated with: the vendor, the users, hospital management, the Norwegian Nurses Association and the research community. To illustrate how the content varied across these arenas, we describe our interventions with two of them, one associated with the vendor and one associated with the users.

The first visit to the vendor was made by the first author in December 2005. He was invited to present results from his study of the HealthSys EPR implementation at UNN in 2004 (documented in Ellingsen & Monteiro, 2005). Prior to this meeting, one of the researchers had presented his findings to senior management at another university hospital. HealthSys's CEO knew about this and had expressed concerns to the researcher that the vendor (and its IS portfolio) had not been treated 'fairly', risking the loss of a large contract with that hospital. In a subsequent meeting in the vendor's offices, ten members of the senior management team discussed the issue again with the researcher. A major aim for the vendor was, as the CEO put it, to "put the facts right". The researcher struggled to explain that this kind of (ethnographic) research represented a specific perspective and an interpretation (Dourish, 2006). After two hours of discussion, the researcher realised that the participants were still rather reserved and did not find his contributions particularly relevant. In order to improve the situation, the researcher suggested presenting some findings from the nursing plan project at UNN (documented in this paper). Over the next two hours, the researcher increasingly won the participants' attention, especially as he managed to translate his findings into concrete design suggestions. This enabled the vendor's developers to reflect on how to implement changes to the nursing module. A lively and constructive discussion followed.

Specifically, the researcher described how the users tinkered with the global NANDA and NIC classification systems. The users had localised the codes by adding comments in the amendment fields; they had also registered the same code in several categories, arranged the codes in a specific order, referred to them in the reports, omitted them from the plan, etc. This not only ensured the use of the global NIC and NANDA standards, it also ensured a system that was carefully tailored to the particular work practice and to the users' own purposes. The researcher suggested supporting this localised use by specific design suggestions. This included the highlighting of important diagnoses, supporting the reuse of frequently used information and making maintenance of codes easier.

At a later stage, the vendor and the researchers agreed to strengthen the collaboration, as the vendor was in an early phase of building a completely new version of the nursing module. In this process, HealthSys wanted the researchers to interact with the design and implementation process since the company (due to its extensive growth in the health market) faced serious challenges in preserving their long tradition of involving users in the design process. In this way, the vendor was responding to the challenge of scale recognised in projects moving from a local to a global scope (Braa et al., 2004).

Initially, the user arena was easier both to establish and to operate in than the vendor arena. The Department of Special Psychiatry had worked with the nursing module for some time and, despite their expressed satisfaction with the module, they struggled with the functionality of the system. This made our research approach valuable to them for two reasons. Firstly, they perceived us as a relevant discussant. For instance, did the quality of their work improve, and did the quality of the nursing documentation improve? Secondly, the users lacked a 'hotline' to the vendor, and they hoped that we could assume a mediating role.

When presenting our results at a departmental workshop and at several occasions for the department's top management of nurses, we focused on the users' *work practices*. This being familiar, it encouraged the users to engage actively in the discussions voicing their expectations and perceptions, to identify what was currently missing and what they hoped to achieve overall through use of the nursing module. As one of our findings, we discussed how the initial objective of replacing the written report with the plan was largely an illusion. The daily written report served several purposes, such as communicating to the other professions and repeating the content of the plan; ultimately it was a prerequisite for the nursing plan as a good plan required effective reports.

In summary, when engaging with the vendor we focussed on the *product* (HealthSys's nursing module), and when engaging with the users we focussed on their *work practice*. In this regard, our design suggestions were not fixed once and for all, but rather served as the starting point for discussions, reflections and negotiated changes. We also experienced that intervening in several arenas, and thus being 'intermediar[ies] between the different actors (Bal & Mastboom, 2005, p. 8) was of increasing value for the stakeholders. For instance, the users knew that we were in communication with the vendor, so they requested that we mediate their concerns. Similarly, HealthSys knew that we operated frequently in several arenas, and our contributions in the vendor arenas were very much appreciated by the company. We believe that our engagement in these arenas enabled us to 'influence the project's course' (Bal & Mastboom, 2005, p. 7) and accordingly, create a *sustainable* network of action (Braa et al., 2004).

# 6 Conclusion

The objective of standardisation has been, and surely will continue to be, changing. The boundary between qualitative phenomena and their (attempted) quantification has historically been an evolving, but contested one. For instance, the quantification (thus standardisation) of initially qualitative phenomena such as temperature, time and music came about in the Middle Ages through socio-technical negotiation processes (Crosby, 1997). It is evident that the historical expansion of the scope of quantification has been met with fierce opposition or, at times, violence (Scott, 1998).

Closer to the immediate topic of this paper, a similar and more recent trend may be observed within the key standardisation institutions such as the International Standardisation Organisation (ISO). From an earlier focus on standardisation of artefacts and products (for example regarding size, technical performance and interoperability), ISO is increasingly involved in the standardisation of previously qualitative issues such as quality (ISO 9000), environmental management (ISO 14000) and social accountability (SA 9000).

The standardisation of service work, which has been our empirical focus, should be recognised against backdrop of these broader trends towards greater standardisation and quantification of non-standard or qualitative 'entities'. With this increased presence of standardisation and quantification comes the increased importance of developing conceptual frameworks to analyse the dynamics of IS-embedded standardisation initiatives.

Standardisation of service work, as exemplified by nursing, provides a particularly valuable platform as it confronts the inherent tension of attempting to achieve improved quality of care while simultaneously enjoying efficiency gains.

The perspective we have developed identifies key mechanisms of transformative, co-constructive practices that constitute standardised service work. It thus contributes to the analytically based, empirically underpinned, critique of an overly simplistic understanding of what is involved. Simultaneously, as we have demonstrated above, it may function as a platform for interventions. It would be analytically fallacious merely to argue for the unattainability of standardisation by identifying shortcomings in specific cases.

# Appendix

## Acronyms used in the text

| CEN | European Committee for Standardization |
|---|---|
| EPR | Electronic Patient Record |
| ICD | International Classification of Diseases |
| ICNP | International Classification on Nursing Practice |
| ISO | International Organization for Standardization |
| NANDA | North American Nursing Diagnosis Association (Commonly used to refer to the taxonomy itself) |
| NIC | Nursing Interventions Classification |
| NOC | Nursing Outcomes Classification |
| SNOMED | The Systematized Nomenclature of Medicine |
| UNN | University Hospital in Northern Norway |
| WHO | World Health Organisation |